\documentclass[twoside]{ilcws10}
\usepackage[latin1]{inputenc}
\usepackage[dvips]{graphicx,epsfig,color}
\usepackage{wrapfig,rotating}
\usepackage{amssymb,amsmath,array}
\usepackage{wasysym}
\usepackage{cite}                      
\usepackage[small, bf, it]{caption}    

\begin{document}
\title{Cherenkov Detector Prototype \& Testbeam 2009}
%
\author{Christoph Bartels$^{1,2}$, 
  Anthony Hartin$^{1}$, 
  Christian Helebrant$^{1,2}$, 
  Daniela K\"afer$^1$, and 
  Jenny List$^1$
  \thanks{The authors acknowledge the support by DFG Li 1560/1-1.}
  \vspace{.3cm}\\
  1- Deutsches Elektronen Synchrotron (DESY) - Hamburg, Germany \\[0.4mm]
  2- Universit\"at Hamburg, Germany          \\[-1.4mm]
}

\maketitle

\begin{abstract}
  Precise knowledge of all beam parameters is crucial to fully exploit 
  the physics potential of the International Linear Collider (ILC).
  A sufficiently accurate measurement of the beam polarisation can only 
  be achieved using dedicated high energy Compton polarimeters combined 
  with well-designed arrays of Cherenkov detectors.
  This note focuses on the design and detailed simulation of a suitable 
  Cherenkov detector prototype and provides an overview of first results 
  from a highly successful beam test period.
\end{abstract}

\section{High energy polarimetry at the ILC}
At the ILC, some beam parameters have to be measured at a permille level precision to 
fully exploit the physics potential of machine and detectors~\cite{bib:RDR,bib:POWER}.
However, contrary to luminosity and beam energy measurements for which similar precisions 
were already achieved at previous colliders, this accuracy is unprecedented for measurements 
of the beam polarisation.

The polarisation measurement at the ILC will combine the measurements of two dedicated 
polarimeters, located upstream and downstream of the $e^+e^-$ interaction point, and data 
from the $e^+e^-$ annihilations themselves.
While $e^+e^-$ annihilation data will finally provide an absolute scale, 
the polarimeters allow for fast measurements, can give feedback to the machine, 
reduce systematic uncertainties and add redundancy to the system~\cite{bib:EP-paper}.

\begin{figure}[h!]
  \begin{picture}(14.0, 5.5)
    \put( 0.00, 0.00)  {\epsfig{file=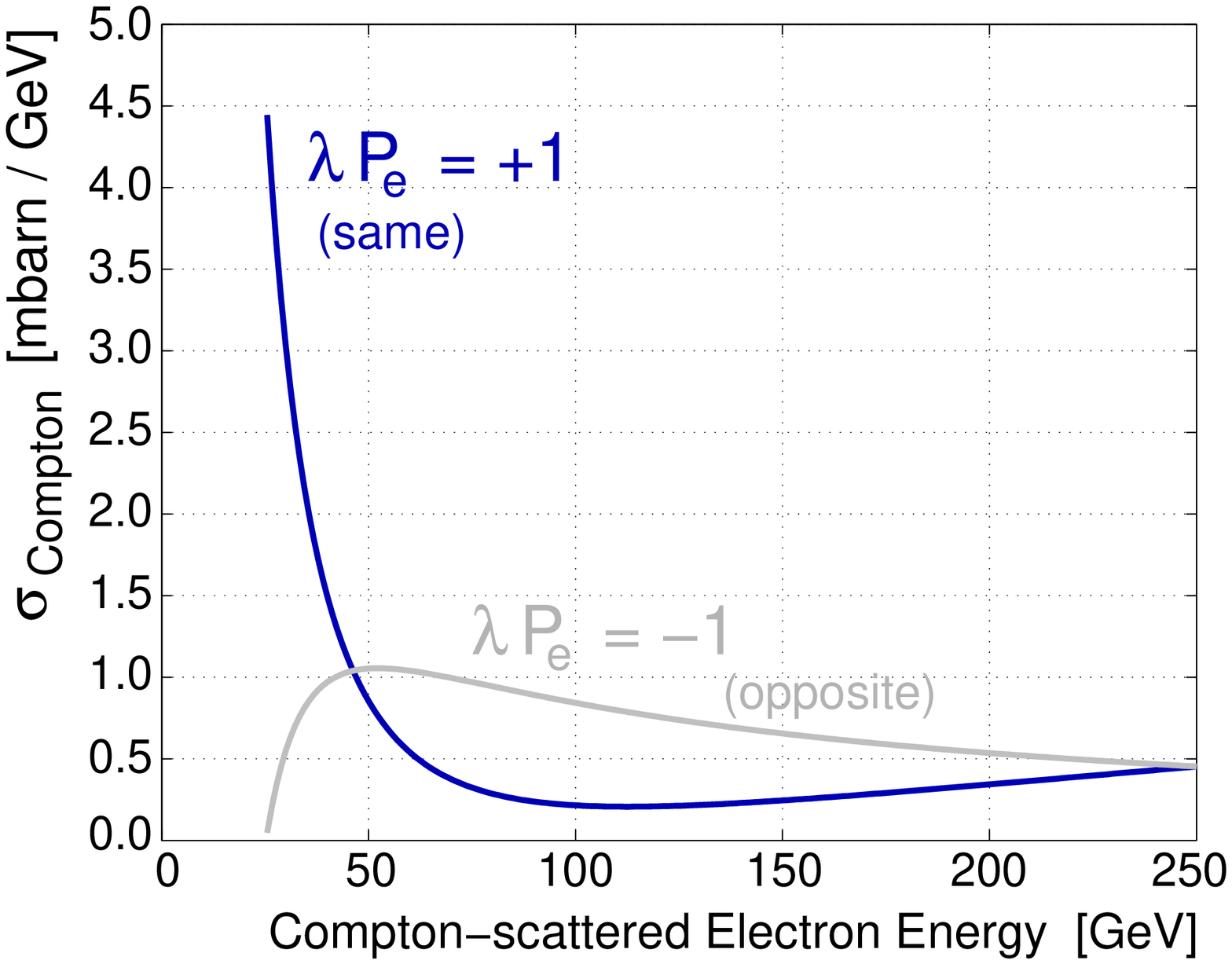, bb=22 175 553 590, clip= , width=0.50\linewidth}}
    \put( 7.10, 0.00)  {\epsfig{file=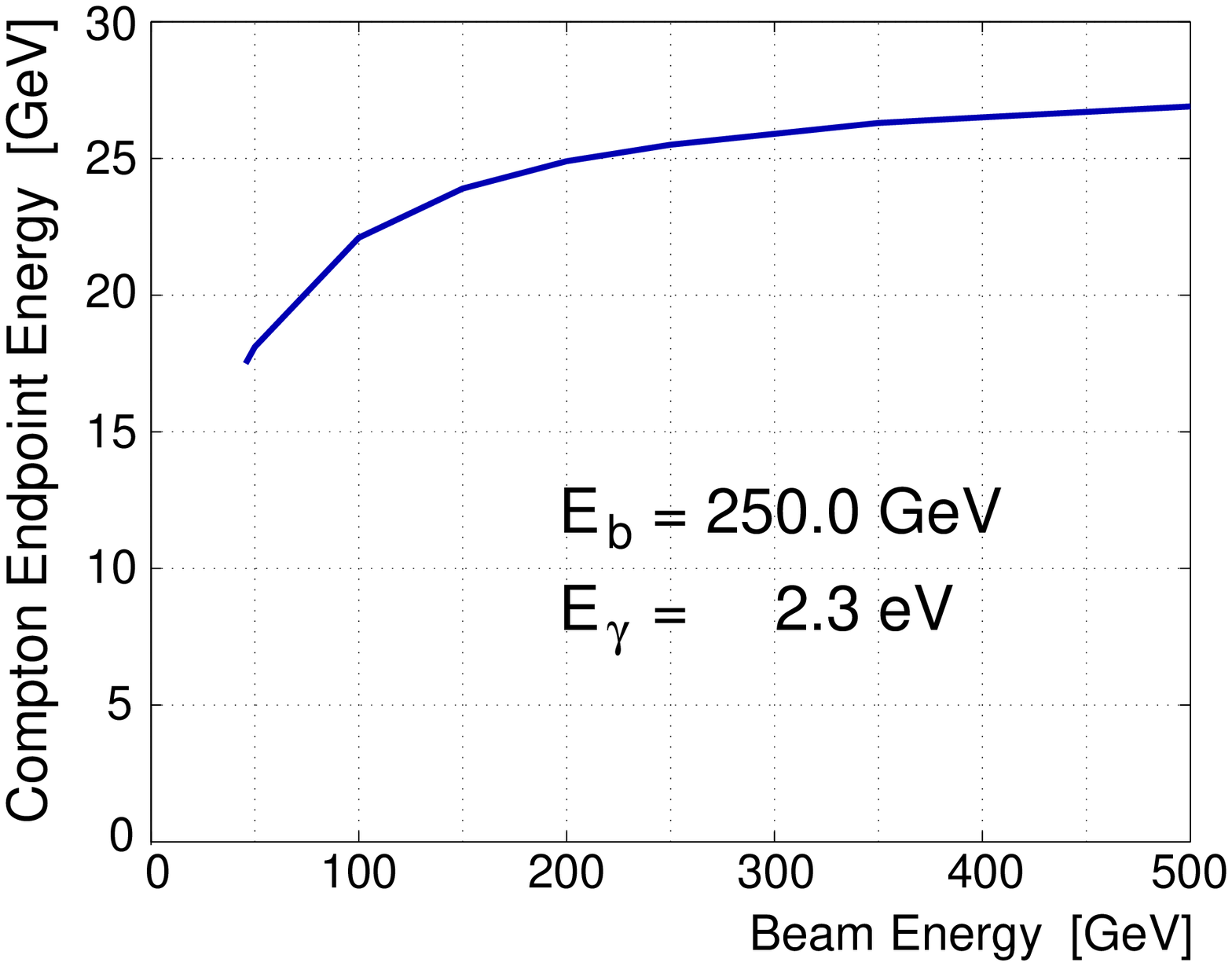, bb=22 175 553 590, clip= , width=0.50\linewidth}}
    \put( 0.00, 0.00)  {(a)}
    \put( 7.20, 0.00)  {(b)}
  \end{picture}
  \caption{
    (a) Compton differential cross section vs.\ scattered electron energy 
    for same (dark curve) and opposite (light curve) helicity configuration 
    of laser photon and beam electron. 
    (b) Compton edge energy dependence on the beam energy.}
  \label{fig:ComptonProc} 
\end{figure}

Both upstream and downstream polarimeters will use Compton scattering of high power 
lasers with the electron and positron beams~\cite{bib:RDR, bib:EP-paper}. 
Circularly polarised laser light hits the particle bunches under a small angle and 
typically in the order of 1000 electrons are scattered per bunch. The energy spectrum 
of the scattered particles depends on the product of laser and beam polarisations, 
so that the measured rate asymmetry w.r.t.\ the (known) laser helicity is directly 
proportional to the beam polarisation. 
Figure~\ref{fig:ComptonProc}(a) shows the Compton cross section versus scattered electron 
energy for a beam energy of $E_b=250$~GeV and a photon energy of $E_{\gamma} = 2.3$~eV. 
The large polarisation asymmetry is clearly visible near 25.2~GeV, the Compton edge energy. 
In addition, this edge energy hardly depends on the beam energy as can be seen in 
Figure~\ref{fig:ComptonProc}(b).

\begin{figure}[h!]
  \begin{picture}(14.0, 9.1)
    \put( 0.00,-0.20)  {\epsfig{file=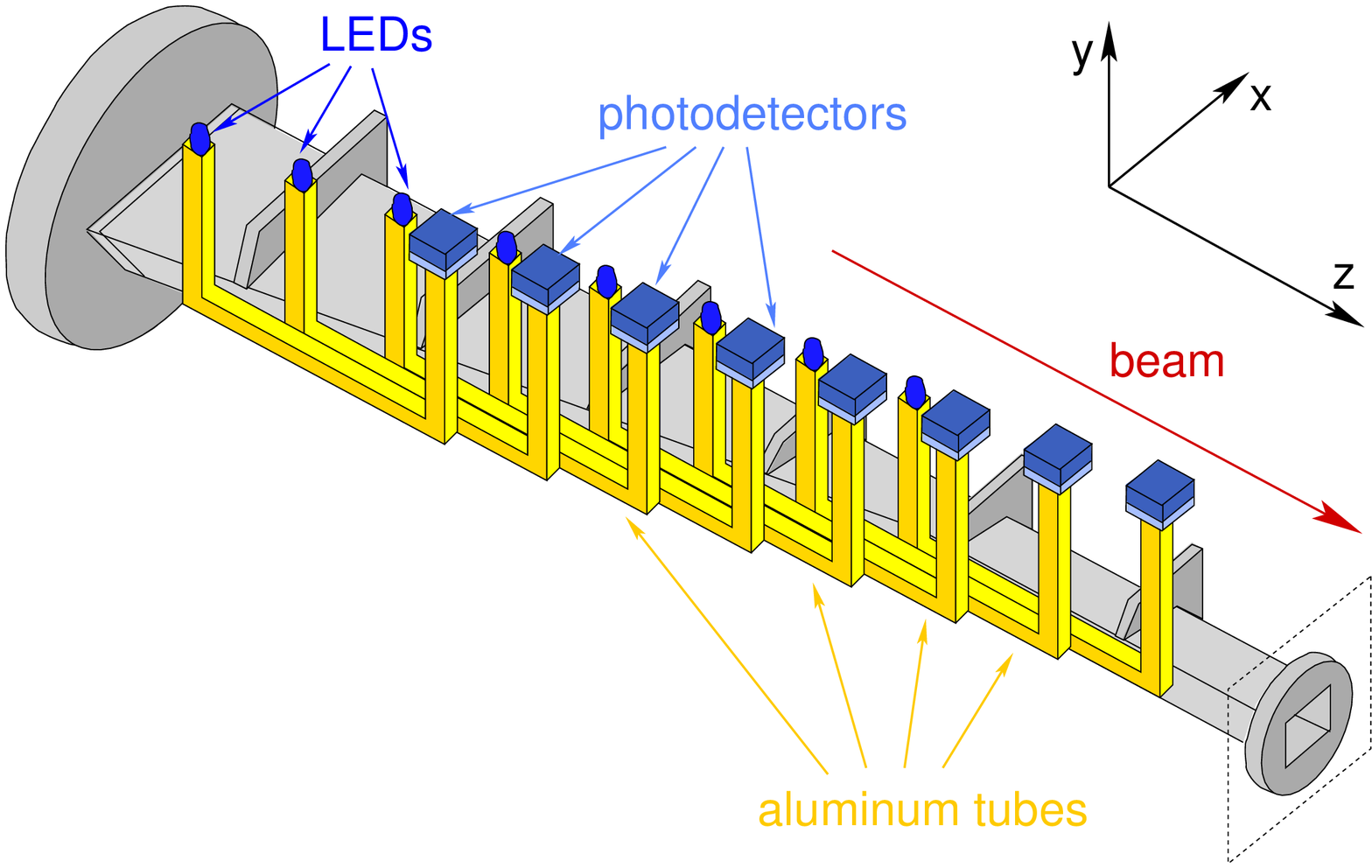, clip= , width=1.00\linewidth}}
    \put(-0.10, 0.00)  {\epsfig{file=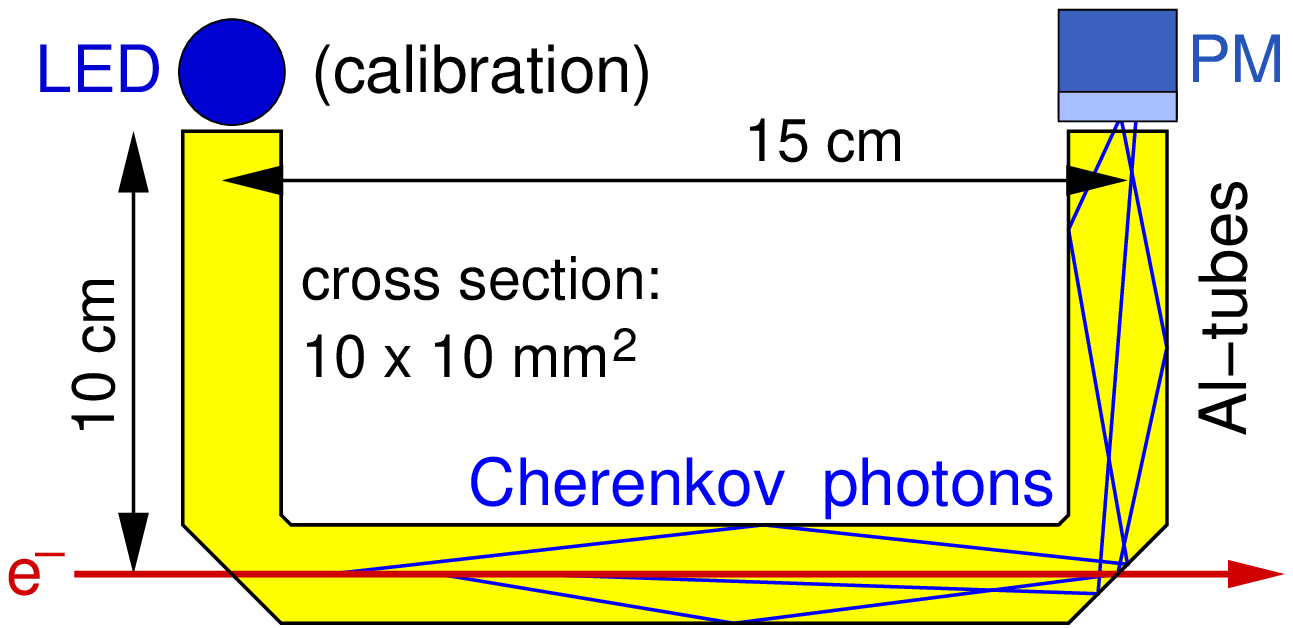, clip= , width=0.45\linewidth}}
    \put( 0.00, 8.50)  {(a)}
    \put( 0.00, 3.50)  {(b)}
  \end{picture}
  \caption{ 
    (a) Illustration of a segmented Cherenkov detector; for better 
    visibility with 8 instead of the forseen 20 readout channels. 
    (b) Sketch of one gas-filled aluminium channel.
  }
  \label{fig:ILDpol-CherTube}
\end{figure}

Since the electrons' scattering angle in the laboratory frame is below $10\;\mu$rad, 
a magnetic spectrometer is used to transform the energy spectrum into a spacial 
distribution and lead the electrons to the polarimeter's Cherenkov detector. 
It will consist of staggered `U-shaped' aluminum tubes lining the tapered exit window 
of the beam pipe as illustrated in Figure~\ref{fig:ILDpol-CherTube}(a). The tubes will be 
filled with a high-threshold Cherenkov gas ($\mbox{C}_4\mbox{F}_{10}$) and are read out by photodetectors (PDs). 
Compton-scattered electrons traversing the tubes' U-bases generate Cherenkov radiation 
which is reflected upwards to the photodetectors~\ref{fig:ILDpol-CherTube}(b).

Developing a Cherenkov detector suitable for achieving the aforementioned precision 
demands improvements in various areas. In order to study the entire experimental setup, 
a two-channel prototype Cherenkov detector has been designed, simulated, constructed, 
and operated successfully in laboratory and beam tests. 

The same right-handed coordinate system (see Figure~\ref{fig:ILDpol-CherTube}) is used 
throughout this note: the beam travels in positive $z$ direction, the $y$-axis points 
upwards, and the $x$-axis to the left when looking in beam direction.
\clearpage

\section{Prototyp design and simulation}
\subsection{Requirements}
Various requirements driving the design of the actual ILC Cherenkov detectors 
have also been taken into account for the prototype detector and its construction:
\renewcommand{\baselinestretch}{1.00}\small
\begin{itemize}
\item  efficient \& homogenous light response to the primary $e^-$ flux 
  (high reflectivity also \\ for short wavelengths; 
  geometry/surfaces to illuminate PD cathode homogeneously)
\item  gas- and light-tightness 
  (control linearity; stabilize response over macroscopic times)
\item  thin inter-channel walls 
  (avoid loosing electrons; avoid background creation)
\item  robustness w.r.t.\ backgrounds 
  (high-threshold gas avoids low-energetic electron\\ background; 
  good layout keeps PDs/calibration source outside the beam plane)
\item  calibration system 
  (monitor response in-situ; indep.\ of beam availability)
\end{itemize}
\renewcommand{\baselinestretch}{1.00}\normalsize
The last two items lead to the idea of U-shaped channels. With increasing length of the 
U-basis more Cherekov light is produced, but the alignment requirements become more 
stringent and additional reflections will decrease the light yield again. Simulations 
show that a length of 15~cm yields a sufficient amount of light while introducing only 
one additional reflection under a glancing angle. 
Contrary to the ILC-like design of staggered channels, the prototype detector consists 
of only two parallel, non-staggered channels (Figure~\ref{fig:SimGEANT}), but it still 
allows to test all relevant aspects of the full detector.

\subsection{Optical simulation with GEANT4}
\begin{figure}[h!]
  \begin{minipage}[c]{0.45\linewidth}
    \caption{
      Event display of the 2-channel prototype simulation: \newline
      The electron beam (red) passes from left to right through the U-basis of 
      the aluminium tubes filled with perfluorobutane, $\mbox{C}_4\mbox{F}_{10}$, 
      and emits Cherenkov photons (green). 
      These are reflected upwards to a photodetector mounted on the hind U-leg.  
      Both channels are separated by a thin foil (light grey). 
      Due to a surrounding gas-filled box, Cherenkov radiation can also be 
      emitted before/after the electron beam enters/exits the aluminium tubes, 
      but it cannot reach the photodetector.
    }
    \label{fig:SimGEANT}
  \end{minipage}\hspace*{2.5mm}
  \begin{minipage}[c]{0.52\linewidth}
    \begin{picture}(8.0, 5.0)
      \put( 0.50,-0.40)  {\epsfig{file=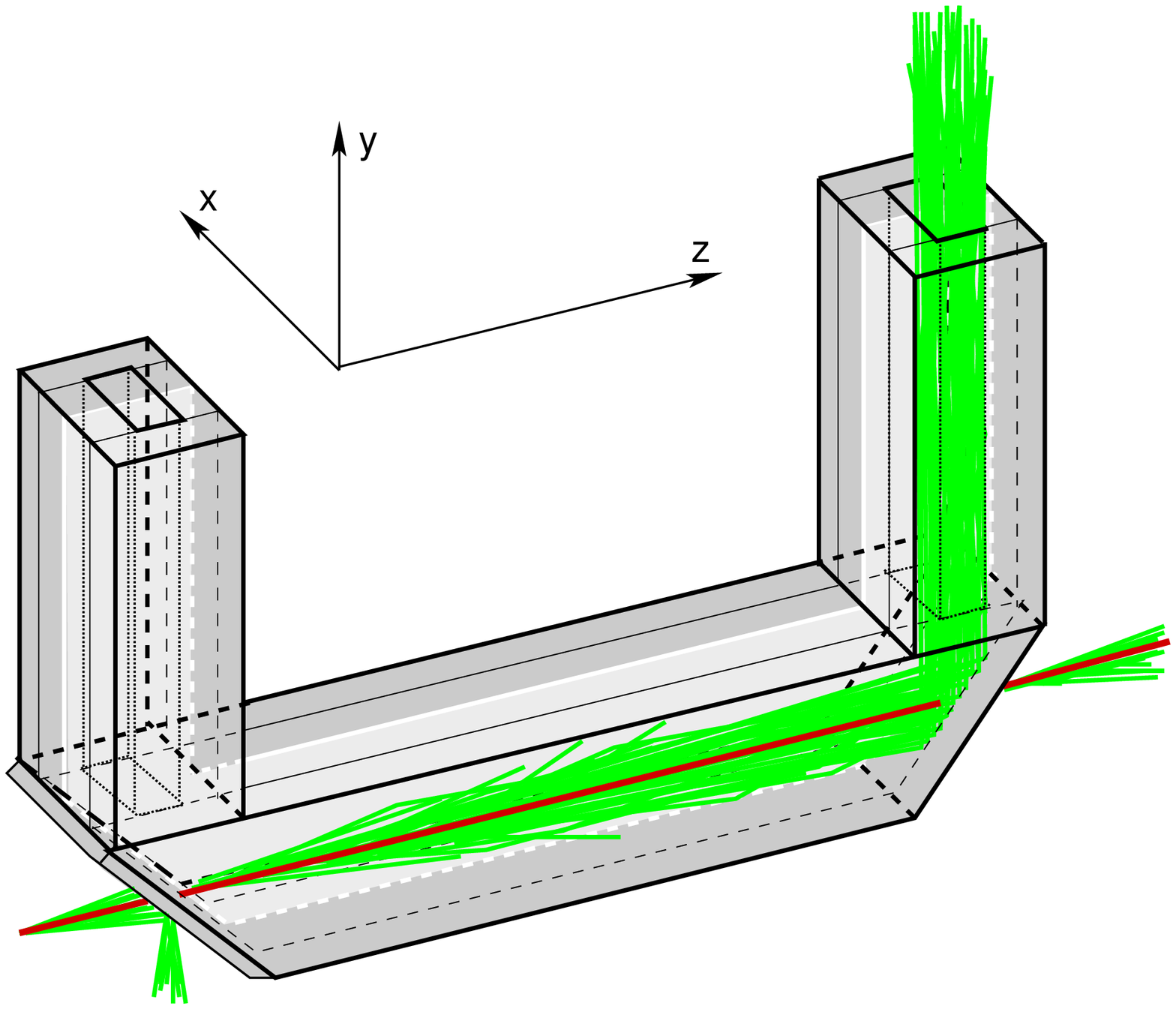, clip= , width=0.92\linewidth}}
    \end{picture}
  \end{minipage}
\end{figure}

For the design of the prototype detector and the interpretation of the testbeam data, 
an optical simulation based on GEANT4~\cite{bib:geant4} has been created. 
The prototype is simulated according to the Technical Drawings and is surrounded by a 
wide box filled with perfluorobutane ($\mbox{C}_4\mbox{F}_{10}$) as Cherenkov gas. Figure~\ref{fig:SimGEANT} 
shows the internal channel structure with the electron beam passing from left to right 
through the U-basis of the right-hand side 
detector channel. 
Although the surrounding base box is not shown, the ambient Cherenkov gas still 
encompasses the entire structure. This leads to Cherenkov radiation emitted outside 
the aluminium tubes, i.e.\ before the beam enters and after it exits the internal 
channel structure. However, light emitted outside the channel volumes cannot reach 
the photodetectors mounted on the hind U-leg. 
Care was taken to properly simulate the inner channel walls by introducing two 
different types of aluminium with average refractive indices of 
$R_{\rm diam} \approx 0.83$ for the diamond-milled aluminium making up 3 of the 4 inner walls, and 
$R_{\rm roll} \approx 0.37$ for the two 0.15~mm-thin rolled foils comprising the middle wall 
between both channels. The wavelength dependence of both refractive indices is implemented 
by interpolating linearly between different wavelengths for which the reflectivities had 
previously been measured.

The purpose of the optical simulation is to determine some key figures such as the 
photon yield/electron, the average number of reflections and possible asymmetry effects 
due to the chosen geometry or utilized materials. This simulation ends at the top of 
the hind U-leg, right before the produced photons would hit the PD cathode. 

\begin{figure}[h!]
  \begin{picture}(14.0, 6.5)
    \put(-0.05, 0.00)  {\epsfig{file=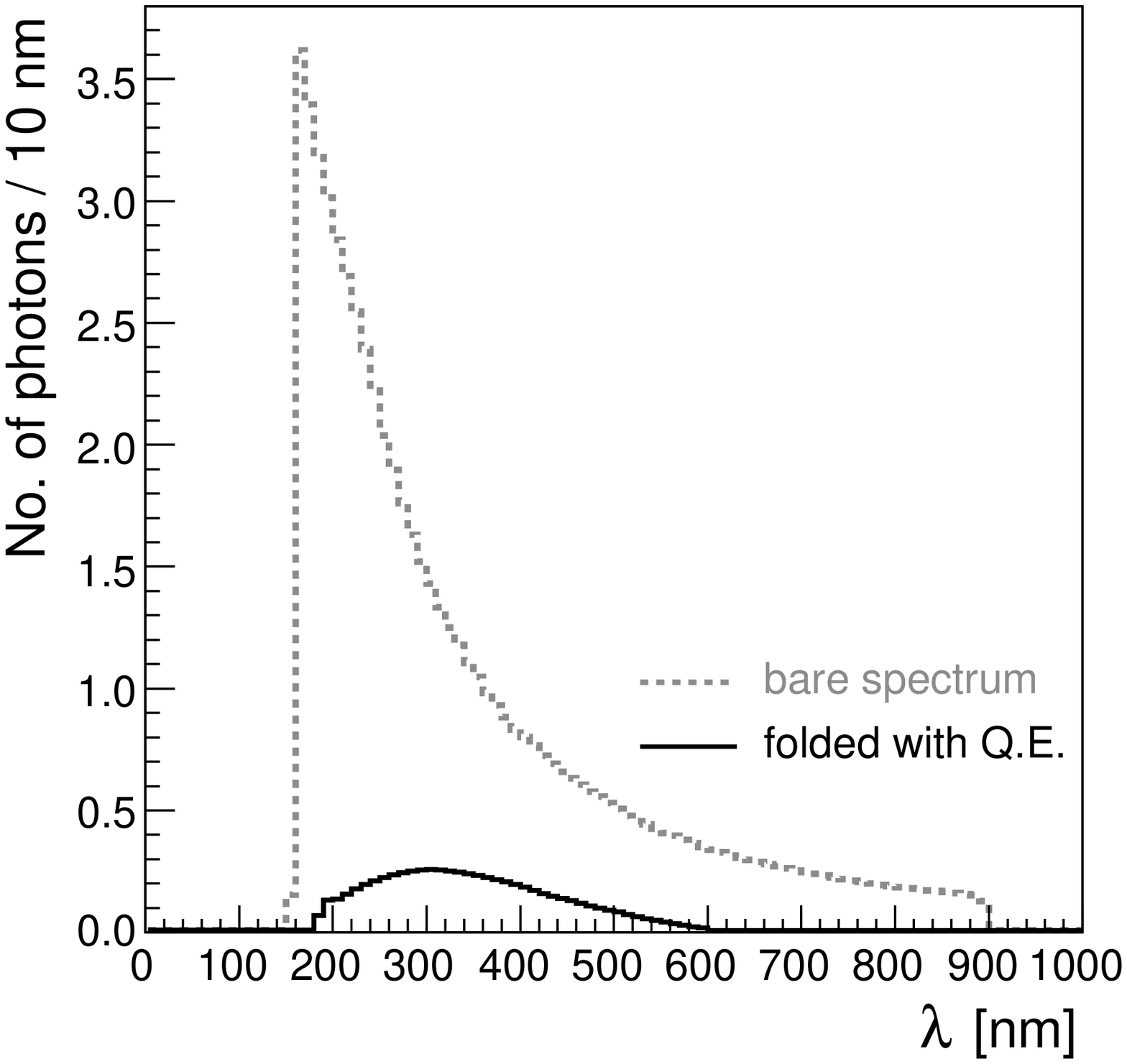, clip= , width=0.50\linewidth}}
    \put( 3.10, 2.90)  {\epsfig{file=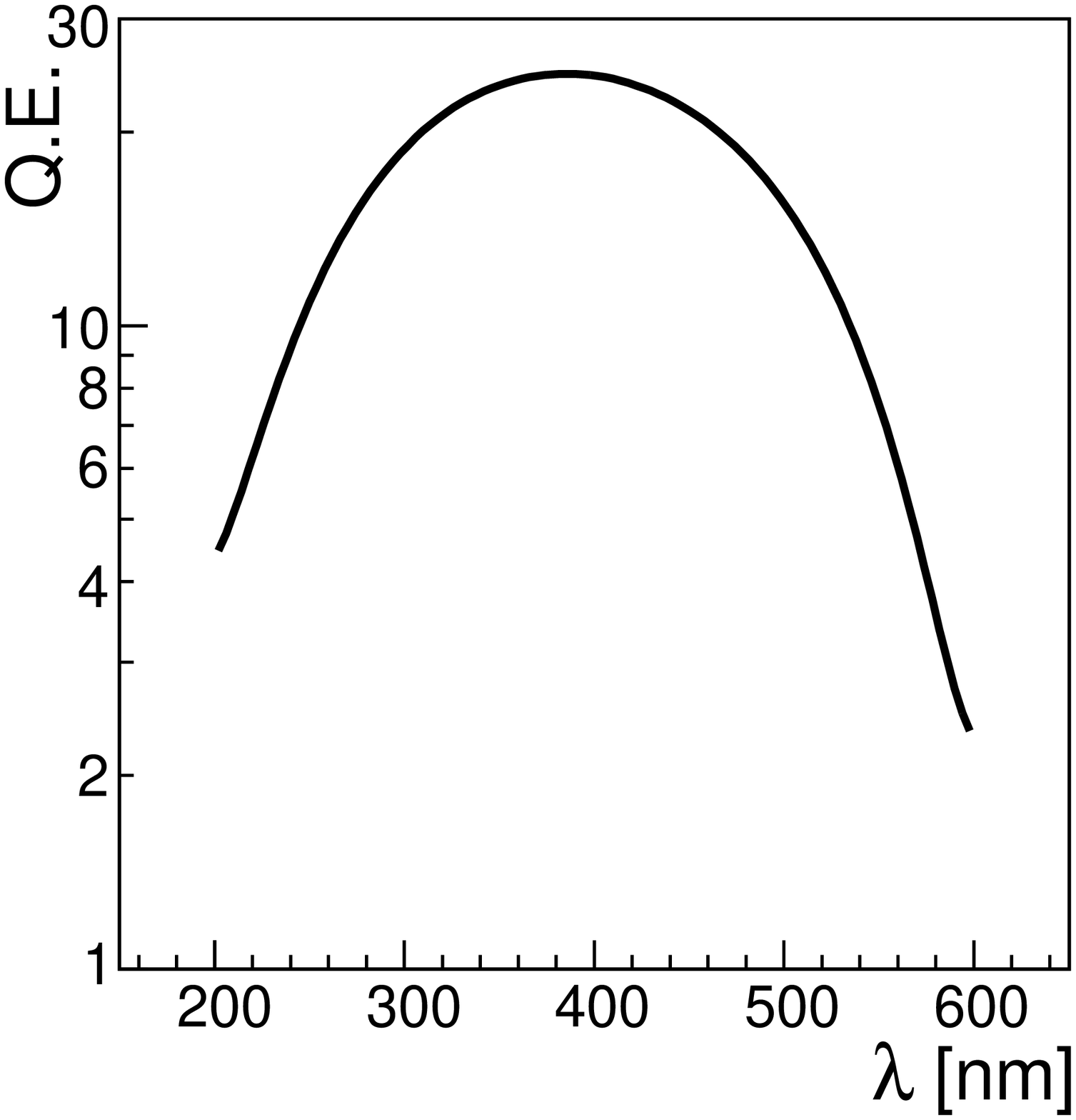, clip= , width=0.25\linewidth}}
    \put( 6.90, 0.00)  {\epsfig{file=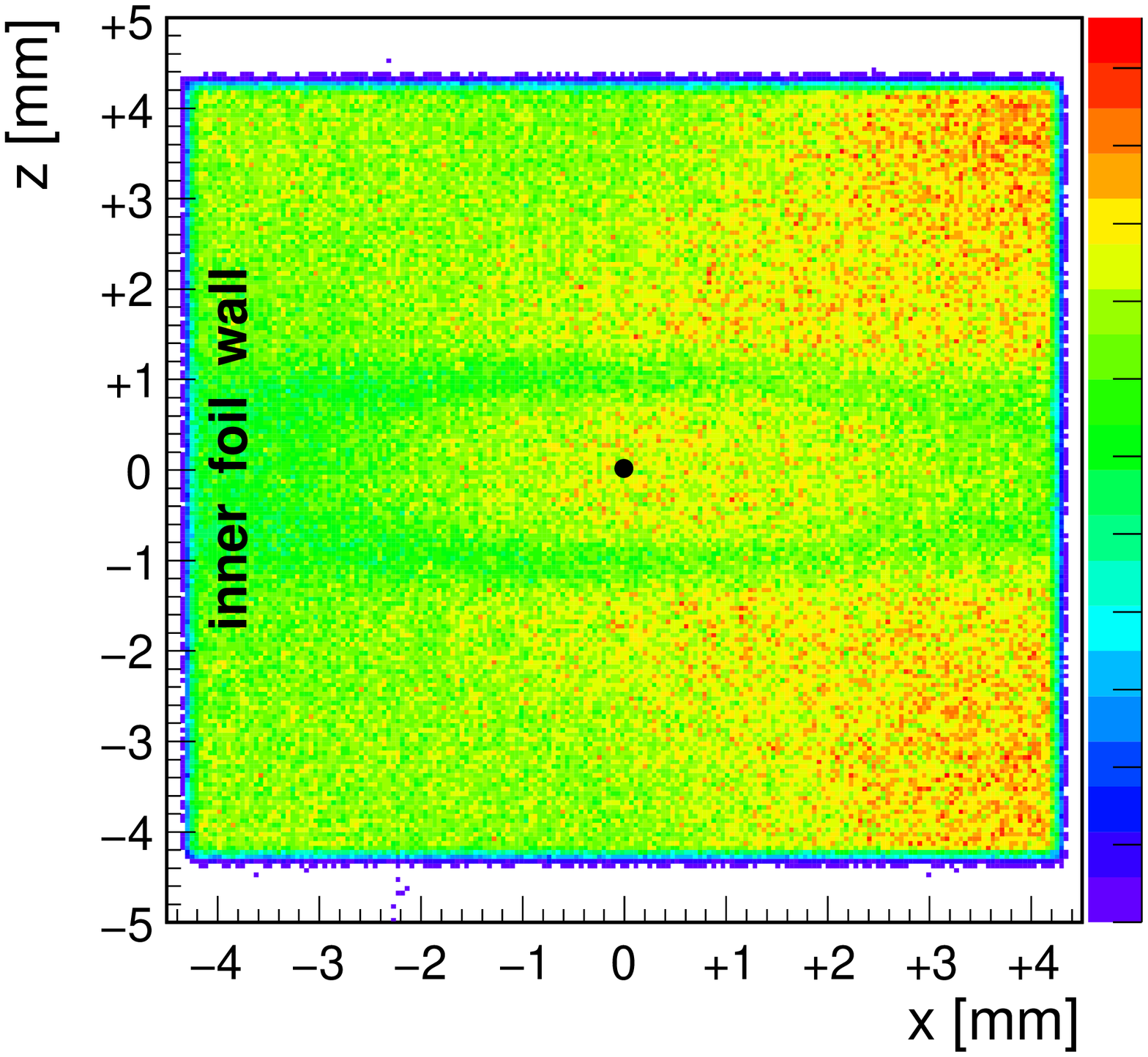, clip= , width=0.505\linewidth}}
    \put( 0.00, 0.00)  {(a)}
    \put( 7.20, 0.00)  {(b)}
  \end{picture}
  \caption{
    (a) Cherenkov spectrum at the PD surface after the optical simulation (dashed grey line) 
    and convoluted with a typical quantum efficiency (solid black line) shown in the inlet. 
    (b) Simulated light yield on the PD cathode for a reduced reflectivity of the inner 
    foil wall ($x=-4$~mm); the black dot indicates the beam entry point.}
  \label{fig:CherSpec-LightYieldSim}
\end{figure}

Figure~\ref{fig:CherSpec-LightYieldSim}(a) shows the wavelength spectrum of these photons 
with its typical $1/\lambda^2$ dependence directly after the optical simulation (dashed grey line) 
and convoluted with a typical quantum efficiency (solid black line).  The applied quantum 
efficiency is shown in the inlet in Figure~\ref{fig:CherSpec-LightYieldSim}(a) and belongs 
to the 2$\times$2 multi-anode photomultiplier (R7600U-03-M4) from {\sc Hamamatsu}~\cite{bib:Hamamatsu}. 
This photodetector is one of two multi-anode photomultipliers (MAPMs) chosen at the very 
beginning of the design phase based on earlier stand-alone studies of different photodetector 
types~\cite{bib:Helebrant-IEEE08, bib:Helebrant-NIM08, bib:Helebrant-Diss}. Due to the layout 
and square geometry of these multi-anode photomultipliers, the channel cross section was 
chosen to be 8.5$\times$8.5~mm$^2$.

To study the influence of the detector geometry and the reduced reflection capability 
of the inner foil wall on the expected light yield on the PD cathode, two scenarios 
are simulated with the same beam parameters (2D gaussian profile, $\sigma_{x} = \sigma_{y} =$ 1.5~mm), 
but different refractive indices for the inner foil wall. 
While the light yield proved to be symmetric about the central $x$ and $z$-axes 
when all four channel surfaces are simulated with the same reflectivity, 
a clear $x$-asymmetry when the inner foil wall is simulated with the correct reduced 
reflectivity, see Figure~\ref{fig:CherSpec-LightYieldSim}(b). 
Overall, the light yield is higher at larger~$x$ and lower at smaller~$x$ allowing to 
conclude that on average the photons are reflected an uneven number of times under a 
glancing angle off the channel walls. 
A simple calculation using a Cherenkov angle of approximately $3^\circ$ and an average 
light path length of 175~mm (half the Cherenkov length plus the length of the U-leg) 
gives an estimate of one single reflection off the channel walls.
The two narrow bands at $z\approx \pm1$~mm, where the light yield is also visibly reduced 
originate from the $90^\circ$ reflection at the end of the Cherenkov section of a channel. 
They also appeared when all channel walls were simulated with the same reflectivity.

The observed light yield asymmetry has also been quantified in the simulation with 
a grid-scan consisting of 4$\times$4 equidistant points (left-hand side detector channel 
and 10,000 electrons per shot). Due to the geometry of the channel layout, 
the $y$-position of the beam entry point on the channel U-basis translates directly to 
the $z$-position on the anode readout plane of the photodetector. However, the results 
of the simulated scans in the $x$ and $y$ ($z$) directions are not displayed here, but 
are included in Figure~\ref{fig:AsymXZ-data}(a,b) on page~\pageref{fig:AsymXZ-data} 
as a comparison with the respective asymmetries calculated from measured beam test data.

\section{Beam test period in spring 2009}
After the final assembly of the prototype detector, its basic functionality was checked 
in various short-term laboratory tests where the gas- and light-tightness was also confirmed. 
In spring 2009, the prototype was taken to the ELSA electron accelerator 
in Bonn and set up in one of the external beam lines directly behind a dipole magnet 
bending the electron beam towards a dump set into the floor. 
The detector was mounted on its turnable base plate and additionally affixed to a stage 
moveable along the $x$- and $y$-axis. Two angles could only be adjusted approximately: 
$\alpha_x$ defining a tilt about the $x$-axis (in the beam-plane) and 
$\alpha_z$ defining a tilt about the $z$-axis in a plane orthogonal to the beam-axis.
While the latter was easily adjusted using a water-level, the former tilt angle $\alpha_x$ 
was more difficult to adjust, since the detector had to be positioned at an angle of 
about $\alpha_x \approx 7.5^{\circ} ... 7.8^{\circ}$ with respect to the horizontal 
to match the downwards slope of the electron beam line.

ELSA comprises three stages: injector LINACs, a booster synchrotron and the 
stretcher ring with a circumference of 164.4~m which leads to a turn time of 
548~ns for relativistic electrons\cite{bib:Hillert}. 
Although the overall beam structure due to the RF acceleration is fixed 
(274 phase space buckets distributed evenly across the ring), the actual 
number of electron bunches per turn can be ajusted via partial filling. 
In addition, the beam spot can be focused to about 1-2~mm and the extraction 
current is tuneable from approximately 10~pA to slightly above 200~pA. 
Both these features lead to numerous electrons traversing the U-basis of the 
detector channels simultaneously, emitting Cherenkov photons and producing 
large and well pronounced Cherenkov signals.

During the test period for the prototype detector, ELSA was operated in 
{\it booster mode} with the electrons being injected at an energy of 1.2~GeV and 
subsequently accelerated to 2.0~GeV. Since beam extraction is not possible during 
the refill/acc.\ phase, the beam is extracted in intervalls of $4\,$s for every 
$5.1\,$s cycle (ratio $\approx\;$4:1).
Since no specific triggers were employed, the beam clock signal looped through a 
function generator was used to provide the necessary gate for the readout electronics 
(a charge sensitive analog-to-digital converter, or QDC). The gate length was adjusted 
such that the detector integrated over all electron bunches of one complete ELSA turn.

\subsection{First signals and prototype detector alignment}
\begin{figure}[h!]
  \begin{picture}(14.0, 6.2)
    \put(-0.10, 0.00)  {\epsfig{file=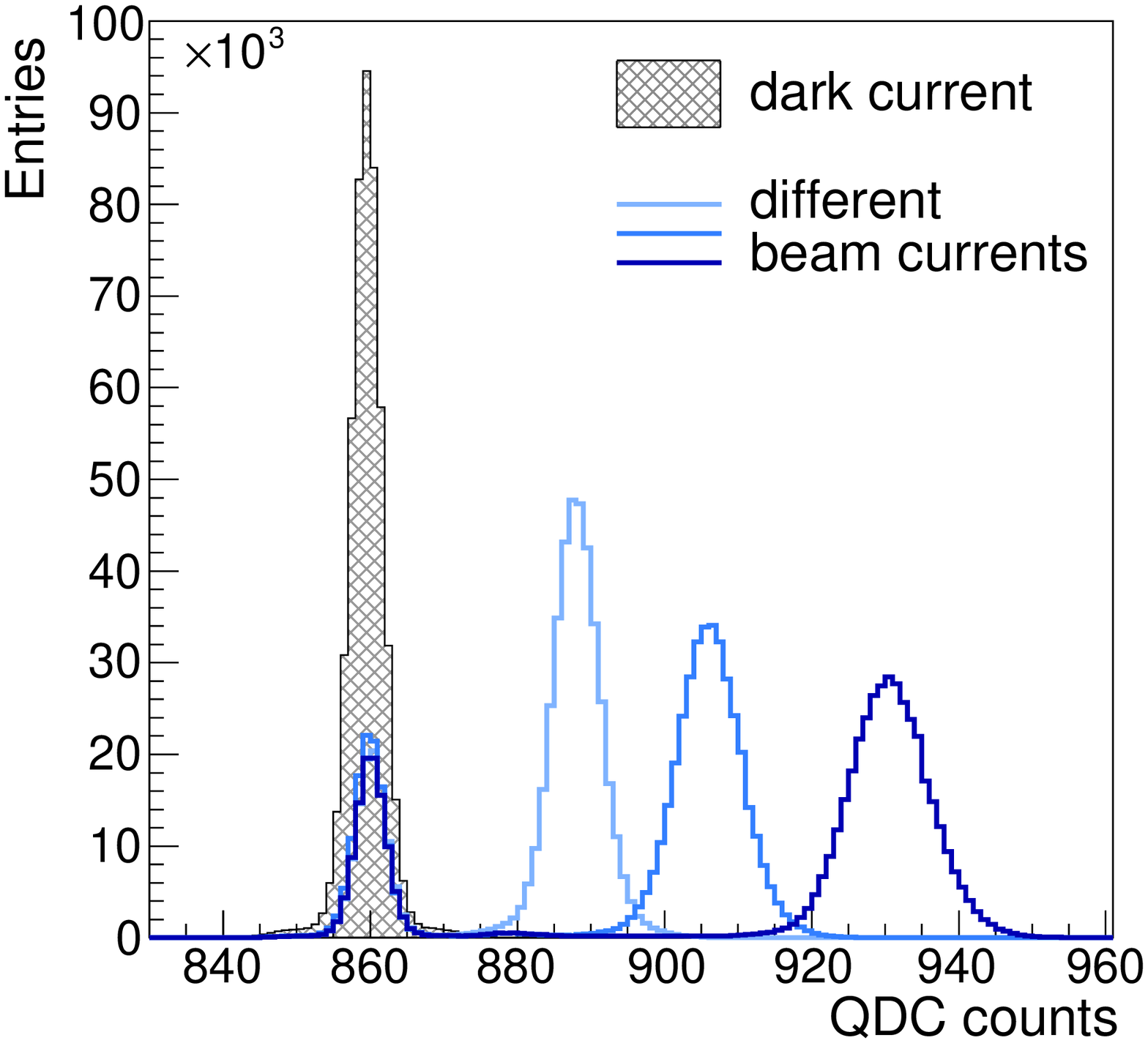, clip= , width=0.50\linewidth}}
    \put( 7.10, 0.00)  {\epsfig{file=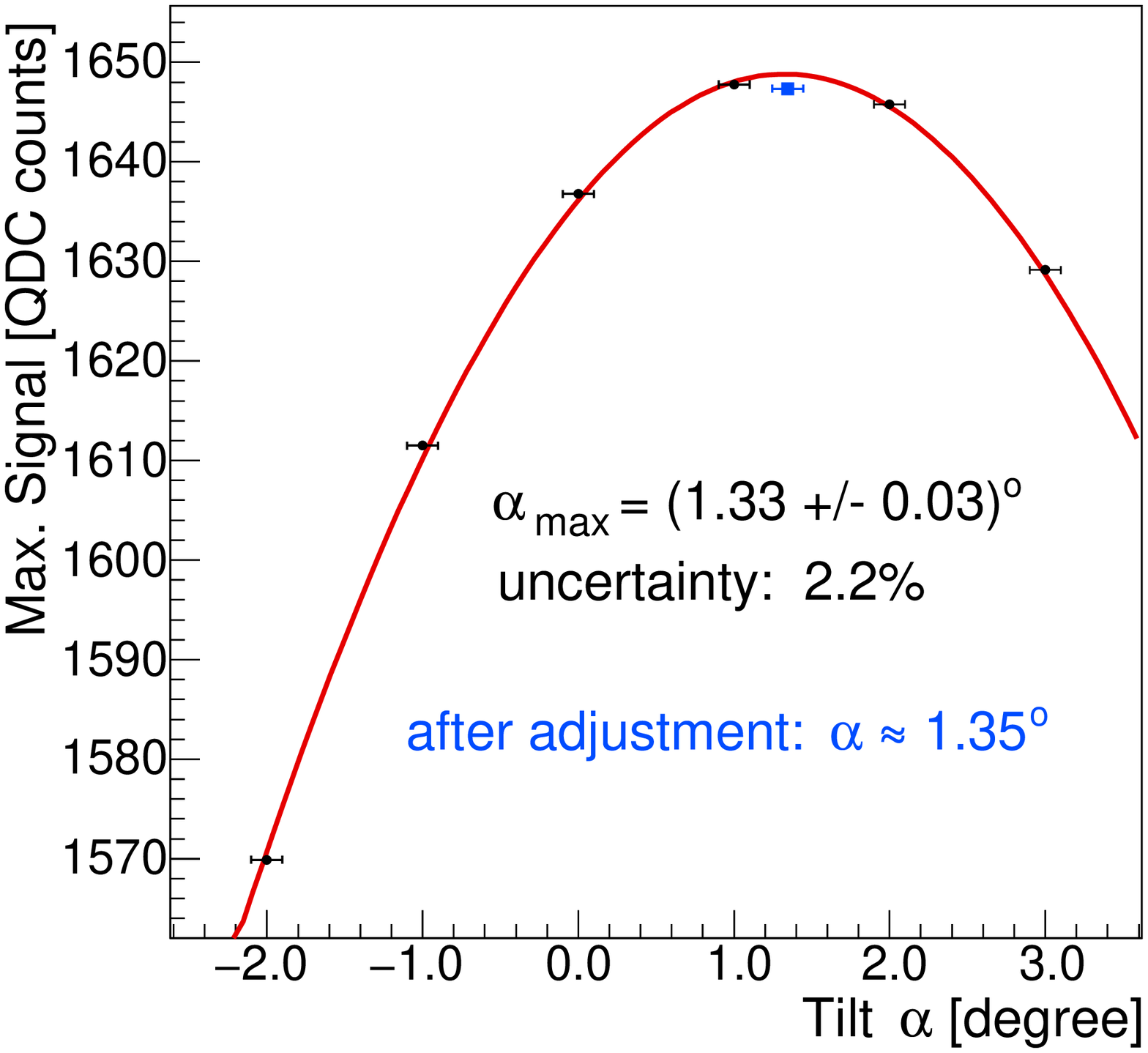, clip= , width=0.50\linewidth}}
    \put( 0.00, 0.00)  {(a)}
    \put( 7.30, 0.00)  {(b)}
  \end{picture}
  \caption{
    (a) The Cherenkov signal (2$\times$2 MAPM) increases with increasing 
    extraction current, while the dark current rate remains stable. 
    (b) The tilt angle $\alpha_y$ is determined from $x$-scans for six 
    different tilt angles (8$\times$8 MAPM). 
    A measurement with adjusted tilt $\alpha_y \approx 1.35^{\circ}$ is also shown.
  }
  \label{fig:M4-DiffBC-TiltScan}
\end{figure}

One of the first measurements performed with beam is shown in 
Figure~\ref{fig:M4-DiffBC-TiltScan}(a). The extraction current was varied from lower 
to higher values and the corresponding Cherenkov signals reproduce this behaviour.
Another important observation is the stability of the dark current (DC) rate. 
The hatched histogram in Figure~\ref{fig:M4-DiffBC-TiltScan}(a) represents the dark 
current (DC) rate recorded without beam, while the open histograms show Cherenkov signals 
for different extraction currents. 
For each open histogram, the area underneath the rightmost peak is about four times 
larger than the area underneath the corresponding dark current peak originating from 
the refill/acc.\ phases without beam extraction (ratio $\approx\;$4:1).

The characteristic DC rate depends primarily on the applied bias voltage and on 
temperature, but not (at least not directly) on the beam current. However, changes 
in beam conditions usually also influence other parameters. Thus, a stable DC rate 
shows that no effects from temperature or beam background are discernible so far. 

The detector position and horizontal tilt are aligned using direct beam data. 
Typically, for a tilted detector the maximal Cherenkov signal at a given beam position 
is smaller than for a perfectly aligned detector.  

While one scan in vertical direction is sufficient to adjust the $y$ position, 
two scans (one on each channel) are preferable. 
To determine the correct tilt in the ($x,z$)-plane, the channel front face is scanned 
horizontally by the electron beam for six different tilt angles $\alpha_y$. 
For all scans, i.e.\ different tilt angles, the highest observed signals are compared 
as shown in Figure~\ref{fig:M4-DiffBC-TiltScan}(b). The best alignment is achieved 
for a tilt angle of $\alpha_y \approx (1.33\pm0.03)^{\circ}$ (represented by the blue square).
This proceedure improves the accuracy on the horizontal tilt angle from 
$\Delta\alpha_y \apprge 3^{\circ}$ to less than 
$\Delta\alpha_y \apprle 0.1^{\circ}$.

\subsection{Further measurements and comparison with simulation}
It is also possible to extract some of the detector dimensions, such as the channel width 
or the distance between channel centers, from beam data. A comparison of these derived values 
and the given prototype specifications can be used to disentangle different effects that are 
either connected to the prototype detector itself, to its alignment w.r.t.\ the beam axis, or 
maybe exclusively to the beam. 

\begin{figure}[h!]
  \begin{picture}(14.0, 6.6)
    \put(-0.10, 0.00)  {\epsfig{file=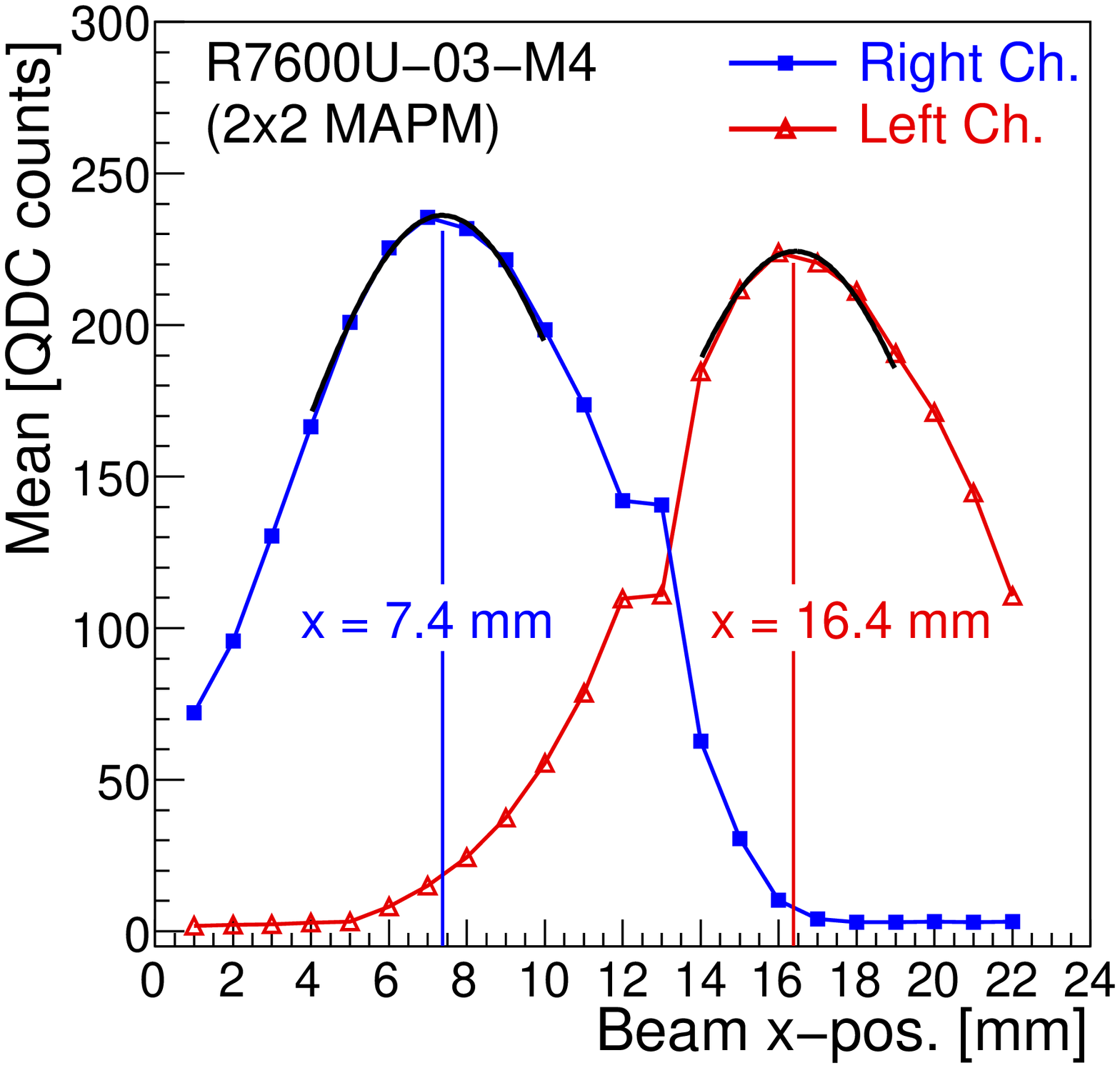, clip= , width=0.50\linewidth}}
    \put( 7.10, 0.00)  {\epsfig{file=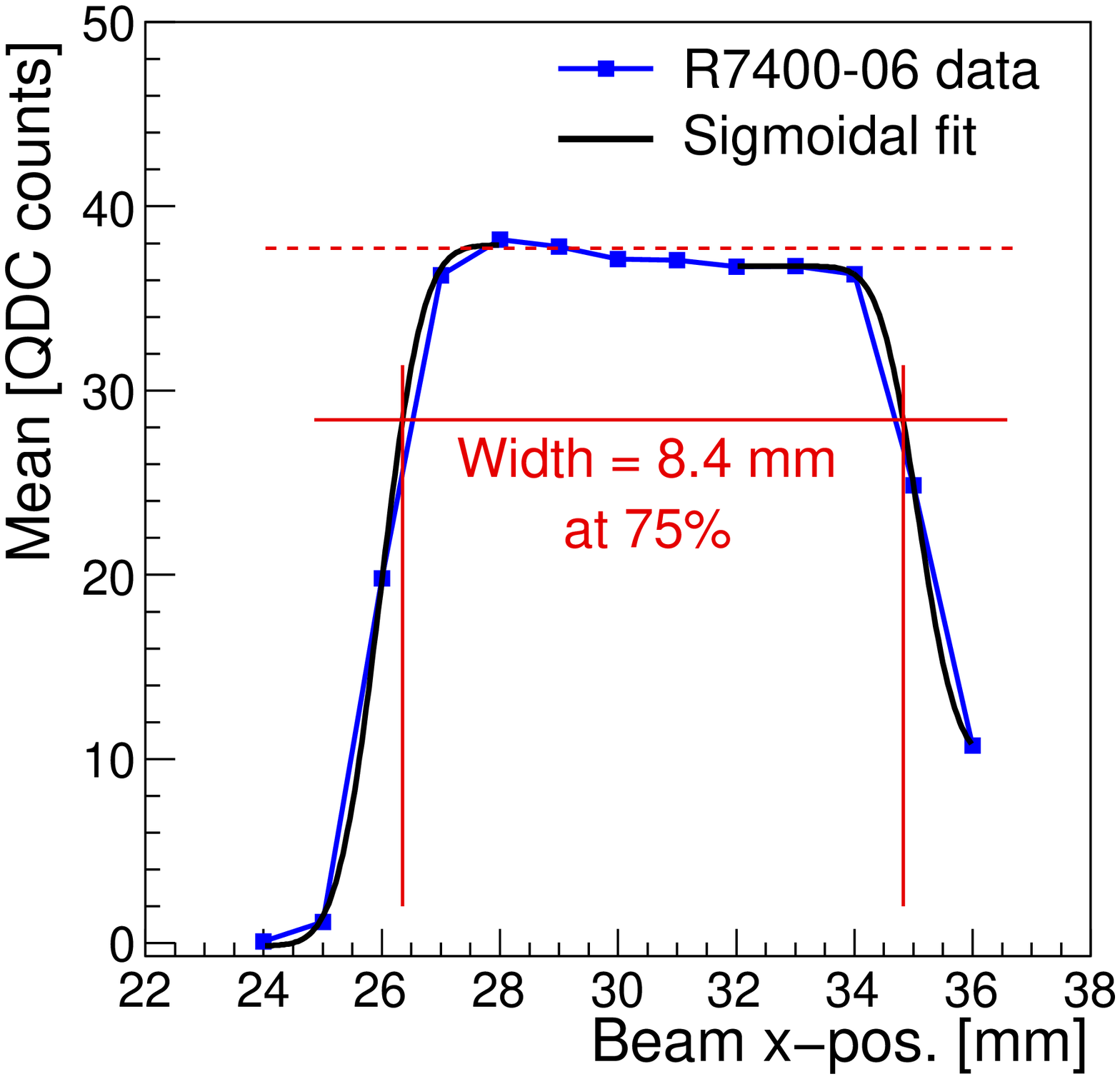, clip= , width=0.50\linewidth}}
    \put( 5.00, 1.15)  {\epsfig{file=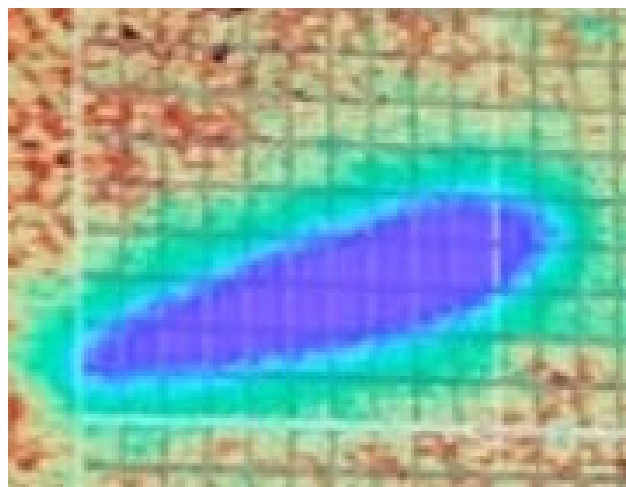, clip= , width=0.09\linewidth, angle=90}}
    \put(10.70, 1.15)  {\epsfig{file=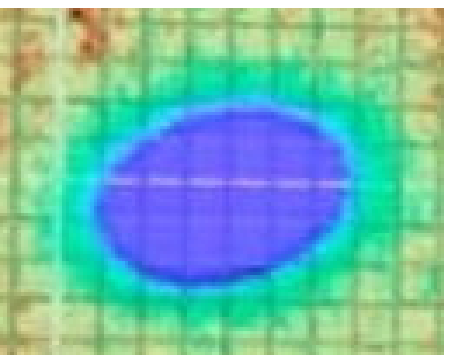, clip= , width=0.09\linewidth, angle=90}}
    \put( 0.00, 0.00)  {(a)}
    \put( 7.30, 0.00)  {(b)}
  \end{picture}
  \caption{
    Beam data $x$-scans, corrected for pedestal contribution:
    (a) 2$\times$2 MAPM and elongated beam spot; 
    (b) SAPM (R7400U-06) and round beam spot. 
    (Abscissae absolute values correspond to different table positions and are meaningless.)
  }
  \label{fig:M4XscanBelong-R7400XscansBround}
\end{figure}

Figure~\ref{fig:M4XscanBelong-R7400XscansBround}(a) shows $x$-scan data recorded with 
the 2$\times$2 MAPM (R7600U-03-M4) for an ellipsoidally elongated beam spot. 
Two gaussian fits indicate the respective channel centres to be at 
$x_{\rm right} = (7.4 \pm 0.1)$~mm and 
$x_{\rm left} = (16.4 \pm 0.1)$~mm, 
leading to a distance of $\Delta x       = (9.0 \pm 0.2)$~mm. 
The nominal distance of  $\Delta x_{nom} = (8.5+0.3)$~mm is given by the sum of the 
channel width and the thickness of the inner foil wall. Both values agree rather well 
considering that effects from different beam spots, the reduced reflectivity of the 
inter-channel wall, or residual misalignment were not yet taken into account.

Figure~\ref{fig:M4XscanBelong-R7400XscansBround}(b) shows the $x$-scan data recorded 
with the R7400U-06 single-anode photomultiplier (SAPM) for a nearly round beam spot. 
From two sigmoidal fits to the edges of the signal and at a 75\%-level of the maximal 
signal strength the effective channel width is measured to be $w = (8.4 \pm 0.2)$~mm. 
Again, the agreement with the nominal width of $w_{nom} = 8.5$~mm is very good, 
indicating that the prototype is functional and behaves as expected. 
The fact that the 2$\times$2 MAPM data in 
Figure~\ref{fig:M4XscanBelong-R7400XscansBround}(a) does not exhibit such a clear 
plateau as the SAPM data is assmued to be due to the different beam profiles during 
the respective measurements. \\

\begin{figure}[h!]
  \begin{picture}(14.0, 5.0)
    \put(-0.10, 0.30)  {\epsfig{file=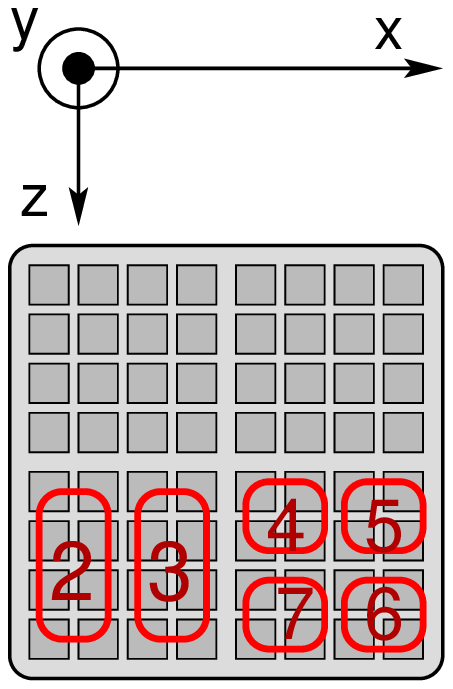, clip= ,  width=0.20\linewidth}}
    \put( 2.90, 0.00)  {\epsfig{file=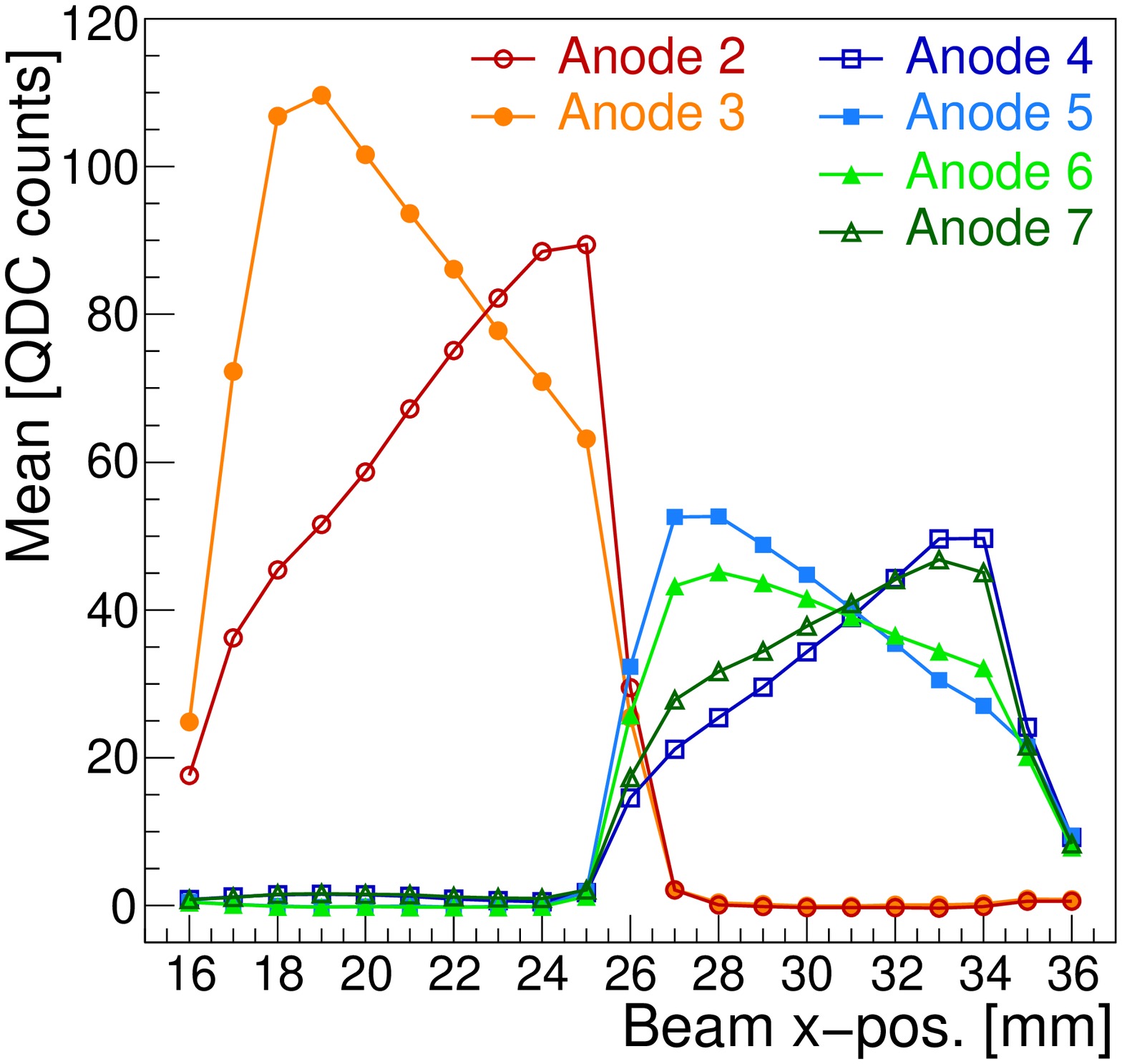, clip= , height=0.37\linewidth}}
    \put( 8.60, 0.00)  {\epsfig{file=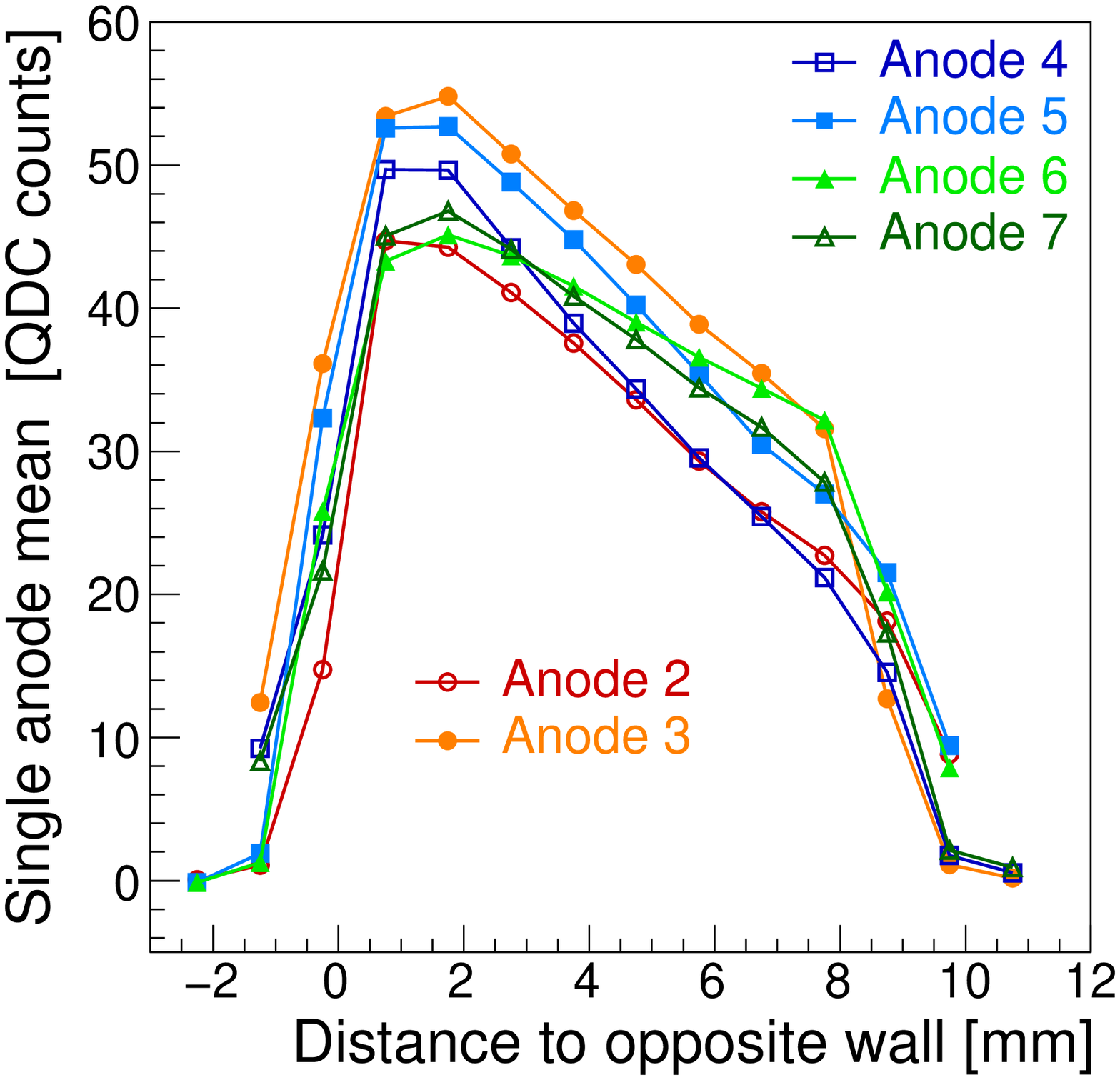, clip= , height=0.37\linewidth}}
    \put( 0.00, 0.00)  {(a)}
    \put( 3.00, 0.00)  {(b)}
    \put( 8.70, 0.00)  {(c)}
  \end{picture}
  \caption{
    Beam data recorded with the 8$\times$8 MAPM: 
    (a) anode readout configuration, 
    (b) $x$-scan across both channels, 
    (c) same data, but accounting for different anode grouping and measuring 
    the distance of the beam entry point w.r.t.\ the opposite channel wall.}
  \label{fig:M64-ac6-XYscan}
\end{figure}

Another measurement series uses the 8$\times$8 MAPM (R7600-00-M64), whose anode is more 
finely segmented and allows studying the intra-channel distribution of the light yield. 
Unfortunately, two out of eight QDC readout channels were broken during the beam test 
period, so that two quadrants of the 8$\times$8 MAPM (i.e.\ 32 anode pads) needed to be 
grouped into the six remaining readout channels.
Figure~\ref{fig:M64-ac6-XYscan} shows 
(a) the chosen anode readout configuration for the 8$\times$8 MAPM, 
(b) data from an $x$-scan across both detector channels versus the beam $x$-position, and 
(c) the same data, but taking into account the different anode groupings for both channels 
and measuring the distance of the beam entry point w.r.t.\ the opposite channel wall.
As can already be seen from Figure~\ref{fig:M64-ac6-XYscan}(b) the highest light yield 
is always measured on the anode pad opposite the respective beam entry point. 
For example, if the beam enters the channel U-basis at the $x$-position of anode~2, 
the highest light yield is measured on anode~3 and vice versa. This can be seen even 
more clearly in Figure~\ref{fig:M64-ac6-XYscan}(c), where all anode responses exhibit 
the same triangular shape. 
The data clearly confirm the simulation result of (on average) one additional reflection 
under a glancing angle off the channel walls apart from the unavoidable $90^{\circ}$ reflection 
from U-basis to U-leg. 

Any remaining amplitude differences still visible in Figure~\ref{fig:M64-ac6-XYscan}(c) 
is either due to different anode sensitivities, residual detector misalignment, 
or a mixture of both effects. \\

\begin{figure}[h!]
  \begin{picture}(14.0, 6.5)
    \put(-0.10, 0.00)  {\epsfig{file=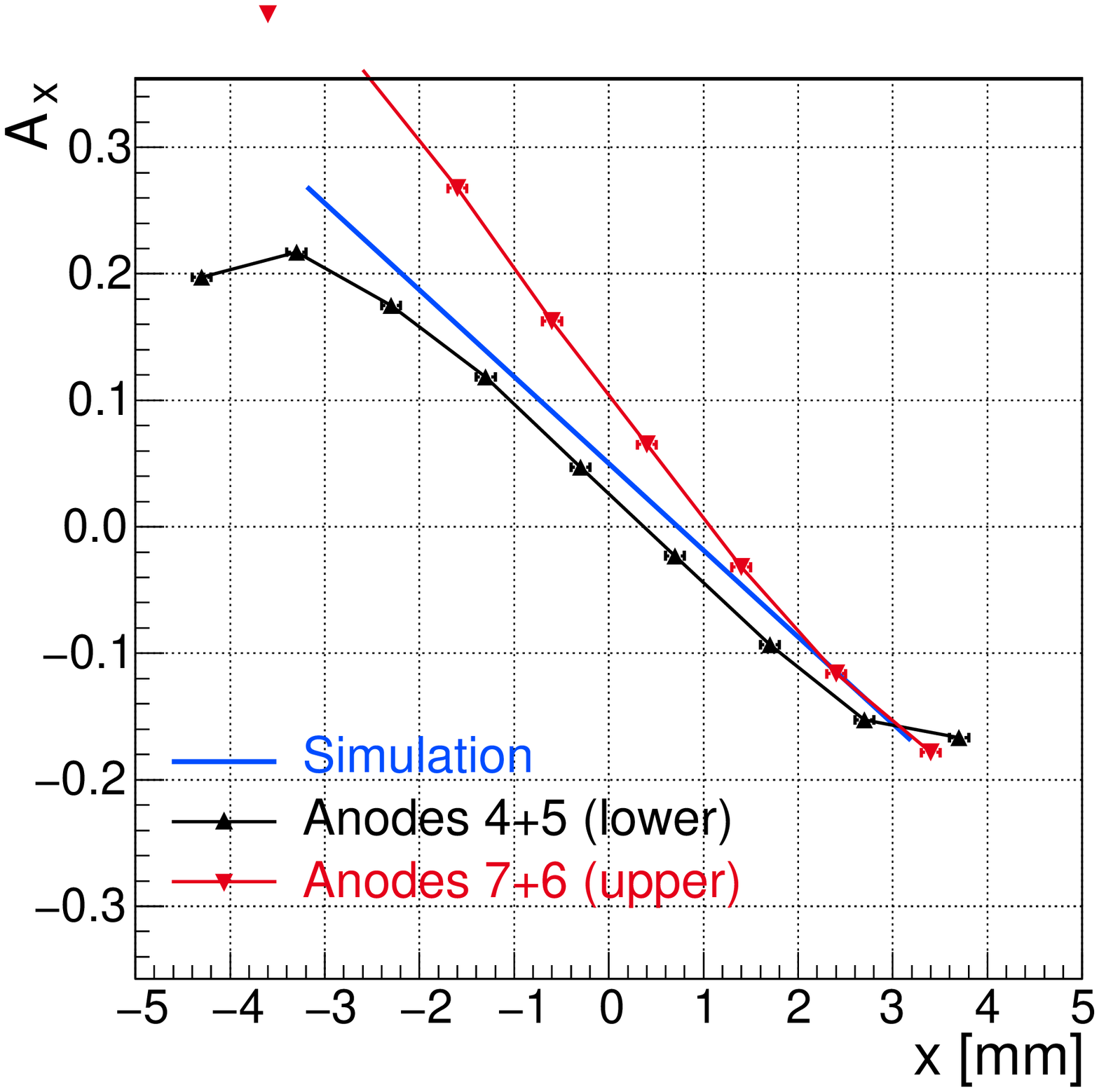, bb=20 30 550 513, clip= , width=0.50\linewidth}}
    \put( 7.10, 0.00)  {\epsfig{file=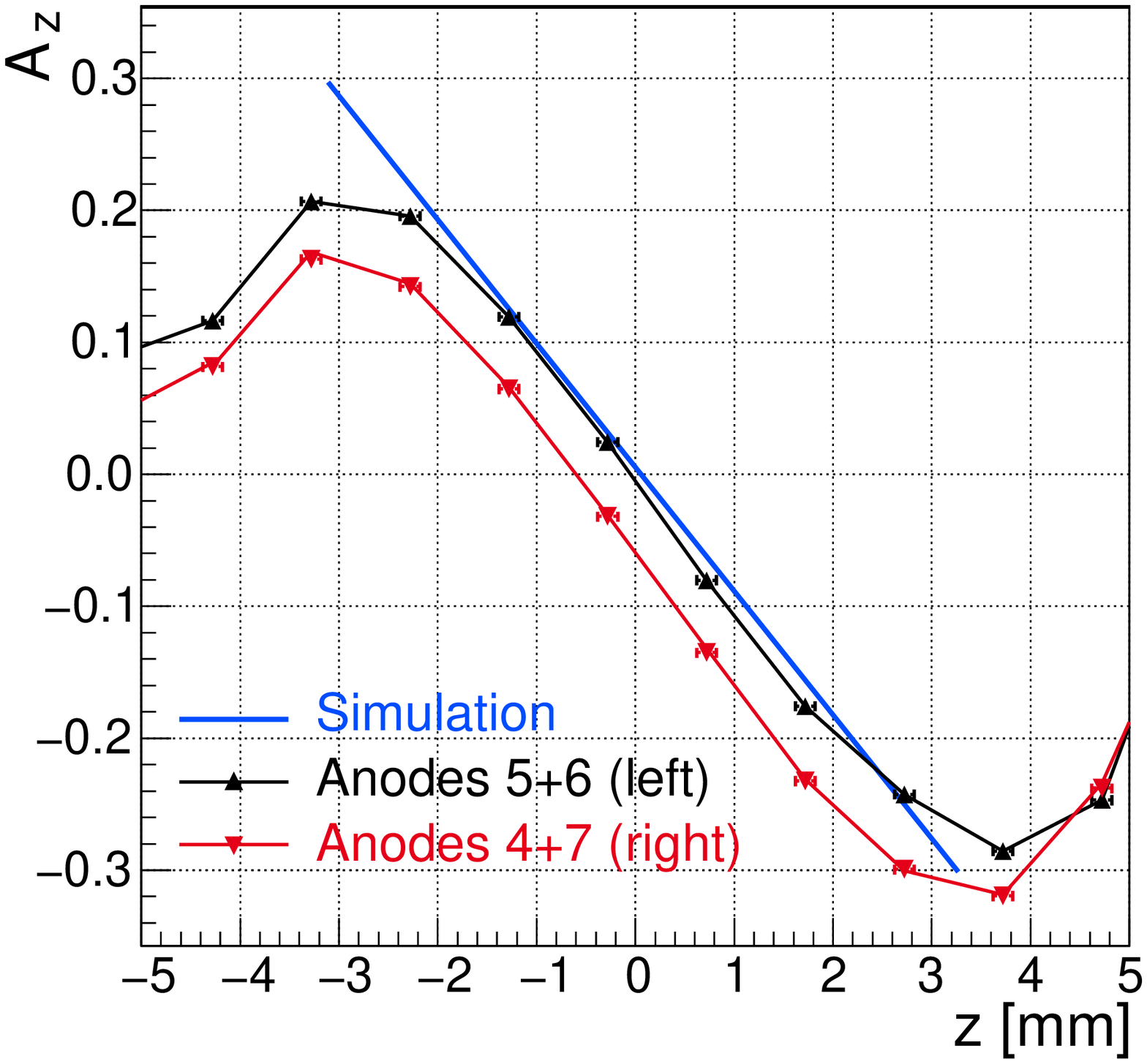, bb=20 30 550 525, clip= , width=0.50\linewidth}}
    \put( 0.00, 0.00)  {(a)}
    \put( 7.30, 0.00)  {(b)}
  \end{picture}
  \caption{
    Asymmetries calculated from testbeam data recorded with 
    the 8$\times$8 MAPM in the above shown anode configuration:
    (a) $\mathcal{A}_x$ from $x$-scan data, and 
    (b) $\mathcal{A}_z$ from $y$-scan data.
  }
  \label{fig:AsymXZ-data}
\end{figure}

The asymmetries in Figure~\ref{fig:AsymXZ-data}(a,b) are derived from $x$ and $y$-scan data 
for the left-hand side detector channel and are directly comparable to the simulation.
For each beam position the corresponding asymmetries $\mathcal{A}_x$ and $\mathcal{A}_z$ 
are calculated by subtracting the light intensity measured in the left (lower) channel 
half from the intensity measured in the right (upper) half:
\begin{displaymath}
  \mathcal{A}_x = \mbox{\LARGE $\frac{ I_x^+ - I_x^- } { I_x^+ + I_x^- }$} \hspace*{10mm} \mbox{ and } \hspace*{10mm} 
  \mathcal{A}_z = \mbox{\LARGE $\frac{ I_z^+ - I_z^- } { I_z^+ + I_z^- }$} 
\end{displaymath}
where $I_x^+$ ($I_z^+$) correspond to the intensities in the left (lower) half of a channel 
and   $I_x^-$ ($I_z^-$) to the intensities in the right (upper) half, respectively.
The non-vanishing value of the simulated $x$-asymmetry at~$x=0$ in 
Figure~\ref{fig:AsymXZ-data}(a) originates from the different reflectivities 
of the inner channel walls (diamond-milled aluminium versus foil reflectivity).

Two $x$-asymmetries are calculated from the beam data: 
one for anodes~4+5 (one for anodes~7+6) 
located   at lower (higher) $z$-values 
atop the     lower (upper)  half of the left detector channel.  
The two $z$-asymmetries are defined for the orthogonal grouping 
of anodes~5+6 (and anodes~4+7) located 
atop the left (right)  part of the channel 
at     higher (lower)  $x$-values, respectively. 
(See Figure~\ref{fig:M64-ac6-XYscan}(a) for details on the readout configuration.)
Both asymmetries, in $x$ and in $y$ direction, are clearly observable in beam data 
demonstrating that the effect is not washed out by imperfections of the channel 
surfaces not modelled in the simulation. This means that also the asymmetries could 
be used for alignment and calibration purposes in the final ILC Cherenkov detectors 
if position sensitive photodetectors are employed. 

Qualitatively the measured asymmetries agree well with those from simulation. 
Especially the $x$-asymmetries are not point-symmetric, but exhibit clear offsets 
in $x$, as expected due to the lower reflectivity of the thin inter-channel wall.
Quantitatively however, some of the measured asymmetries deviate from the 
ideal expectation. The $x$-asymmetry from anodes~7+6 exhibits a different slope, 
while the $z$-asymmetry from anodes~4+7 seems to be shifted. These differences 
hint towards residual misalignment around the $y$ and/or $z$-axes and the influence 
of the non-perfect beam profile also still needs to be taken into account.

\section{Conclusions and  perspectives}
A two-channel prototype Cherenkov detector has been designed, simulated and 
successfully operated under beam conditions. The optical simulation of the 
prototype detector is based on GEANT4 and provides the light distribution 
at the photodetector cathode. 

The modular design of the prototype allows fast exchanges of the photodetector 
and calibration modules and thus facilitates operation during beam tests. 
So far a number of expectations from simulation could already be confirmed 
with a first preliminary analysis of some of the beam data.
One example of a successful application of simulation methods to real data, 
is the extraction of intra-channel position information from various scans.

Further plans include a full analysis of all beam data 
and establishing a permille-level calibration for different photodetectors
applicable also under ILC conditions.

\begin{footnotesize}

\end{footnotesize}

\end{document}